%% file: main.tex
\documentclass[10pt]{article}
\usepackage[utf8]{inputenc}
\usepackage[letterpaper, total={7.5in, 9in}]{geometry}
    \usepackage{graphicx}
    \usepackage{subcaption}
    \usepackage[export]{adjustbox}
    \usepackage{wrapfig}
    \usepackage{float}
    \usepackage[format=plain,labelfont={bf,it},textfont=it]{caption}
\graphicspath{ {./images/} }

\usepackage[numbers,square]{natbib}
\usepackage[scaled]{helvet}
\usepackage[T1]{fontenc}

\usepackage{lineno}

\usepackage{setspace}
\usepackage{mymacros}

\usepackage{authblk}

\usepackage{amsmath}

\usepackage{color,soul}

\usepackage[hidelinks]{hyperref}
\hypersetup{
    colorlinks=false,
    linkcolor=blue,
    filecolor=magenta,      
    urlcolor=blue,
    citecolor=black,
}

\usepackage{algorithm}

\usepackage{algcompatible}
\usepackage{comment}

\title{The amplitude modulation pattern of Gaussian noise is a fingerprint of Gaussianity}
\author[a]{V\'{i}ctor Manuel Hidalgo}
\author[a]{Juan-Carlos Letelier}
\author[b,*]{Javier D\'{i}az}

\affil[a]{Department of Biology, Faculty of Sciences, University of Chile}
\affil[b]{International Institute for Integrative Sleep Medicine, University of Tsukuba}
\affil[*]{Corresponding author: Javier D\'{i}az, diaz.antonio.fn@u.tsukuba.ac.jp  \\ 1-1-1 Tennodai, Tsukuba, Ibaraki, 305-8575, Japan} 

\date{}

\begin{document}

\maketitle

\input{sections/0.Abstract}
\input{sections/1.Intro}
\input{sections/2.Theory}

\input{sections/3.NumericalResults}

\input{sections/4.Discussion}

\section{Acknowledgments}
This work was supported by Fondecyt (ANID Chile) 1210069 and the Japan Society for the Promotion of Science (Grant-in-Aid for Scientific Research (C): 19K12199). We would like to thank Diane Greenstein for editorial assistance.

\bibliography{References.bib}

\end{document}

%% file: sections/0.Abstract.tex
\begin{center}
    \textbf{Abstract}
\end{center}
We introduce a new approach for Gaussianity testing using the envelope of a signal and its coefficient of variation. The envelope of a Gaussian signal follows the Rayleigh distribution, and given that the coefficient of variation of the Rayleigh distribution is invariant, the envelope of Gaussian noise’s coefficient of variation is a universal constant and can be exploited as a discriminating statistic. The main application of this approach is to detect the Gaussianity of time series. However, the coefficient of variation of the envelope is also a measure of amplitude modulation patterns that captures the structure of the Fourier phase profile, making it a useful parameter to differentiate types of non-Gaussianity and for signal classification.

\noindent\textbf{Keywords:}  Envelope, Gaussian noise, Rayleigh fading, Rician fading, Signal detection

%% file: sections/1.Intro.tex
\section{Introduction}
Assessing Gaussianity is a classic problem in signal analysis, and a variety of methods have been devised to infer it from time series. These methods have approached the problem from both the time and frequency domains. While time domain approaches are goodness-of-fit tests (e.g. Jarque–Bera, Kolmogorov-Smirnov)  \citep{thadewald2007jarque,giannakis1994time}, frequency domain methods have been developed based on bispectrums \citep{brockett1988bispectral}. In general, these methods present distinct advantages and drawbacks that make them more or less suitable under different conditions to settle the question of Gaussianity in time series \citep{thode2002testing}. 

In this letter, we present a different approach to Gaussianity testing by exploiting a singular property of Gaussian signals: the envelope of zero-mean Gaussian noise is distributed according to the Rayleigh distribution --- a distribution  with the special property that its coefficient of variation is invariant. Thus, when the envelope of a given time series is analyzed, its coefficient of variation of the envelope (CVE) can be used to test for Gaussianity. 

Here we exemplify how the \cve can be used to infer the Gaussianity of discrete-time signals. \cve is also a measure of amplitude modulation patterns and signal morphology, and Fourier transform phase randomization reveals that \cve is closely connected to the Fourier phase profile of a signal but independent of the spectral profile. Our results show that \cve offers better estimator properties compared to the third and fourth order moments.

%% file: sections/2.Theory.tex
\section{Gaussian coefficient of variation of the envelope}
We consider a signal \s and $\hat{s}(f)$ its Fourier transform (FT). If one chooses to consider only non-negative frequencies  ($f\geq0$), it is possible to represent \s as the complex function \sa, known as the analytic signal, creating an implicit de-facto relationship between the two functions $s(t) = Real(s_a(t)) = A(t)\cdot \cos(\phi(t))$ \citep{picinbono1997instantaneous}.

In this polar representation, $s(t)$ can be written as an amplitude modulation (AM) signal, where $A(t)$ is the \textit{instantaneous amplitude} or \textit{envelope} and $\phi(t)$ is the \textit{instantaneous phase}. These two functions are uniquely defined by the constraint in the frequency domain.  

The analytic signal \sa can be calculated using the original signal and its Hilbert Transform  $s_{a}(t) =  s(t) + i\,\mathcal{H}\{s(t)\}$ \citep{gabor1946theory}. By definition, an analytic signal has no negative frequency components, unlike real-valued signals whose Fourier Transforms are Hermitian functions. This mathematical fact allows \s to be represented as a phasor of variable amplitude, whose envelope function $A(t)$ can be calculated as the norm of the analytic signal $A(t)=\sqrt{s^2(t) + \mathcal{H}\{s(t)\}^2}$. 

The envelope is a non-negative, band limited, and slow-varying signal \citep{ktonas1980instantaneous}. But more importantly, given that the envelope $A(t)$ is always non-negative, then the ratio between its standard deviation and its mean (the statistic known as the coefficient of variation, CV) is always defined and can be used to assess Gaussianity.

A well-known result from telecommunication theory states that a multipath propagation channel with no direct line of sight exhibits Rayleigh fading, and thus its envelope follows a Rayleigh distribution \citep{schwartz1995communication}:
\begin{equation}
    f_{Rayleigh}(x;\sigma)=\frac{x}{\sigma^2}\cdot e^{-x^2/(2\sigma^2)}
\end{equation}
where the variable $\sigma$ is known as the scale parameter.

The core of the demonstration is that if $X$ and $Y$ are two zero-mean Gaussian-distributed random variables with equal variance $\sigma^2$, then the random variable defined by $Z=\sqrt{X^2 + Y^2}$ follows a Rayleigh distribution with scale parameter $\sigma$. This condition is met when using the analytic representation of Gaussian signals because the Hilbert transform is a linear operator, and the Hilbert transform of a Gaussian signal is Gaussian.

The Rayleigh distribution has a special peculiarity. Both the mean and the standard deviation of this distribution contain the scale parameter $\sigma$ \citep{papoulis1989probability}:
\begin{align}
\mu &= \sigma\sqrt{\frac{\pi}{2}}  & 
 std &= \sigma\sqrt{\frac{4-\pi}{2}}   
\end{align}
Thus, the \textit{coefficient of variation} ($CV = std / \mu$) of the Rayleigh distribution is a constant we call $m$:
\begin{equation}
   m = \sqrt{(4-\pi)/\pi} \approx 0.523
\end{equation}

This peculiarity came to our attention while studying the neural oscillations produced by the vertebrate olfactory epithelium. We found that the envelope of these "beta waves" closely fit the Rayleigh distribution and their CVE had a mode close to $m$ \citep{diaz2007amplitude}. Interestingly, we found a similar result reported in the EEG field regarding the human alpha rhythm \citep{berger1929elektroenkephalogramm,adrian1934berger} in which \citet{motokawa1943statistisch} approximated the Rayleigh distribution from human alpha rhythm signals. To our knowledge Motokawa’s report was the first to point out the Gaussianity of neural signals.
We confirmed this result ourselves \citep{hidalgo2021envelope} and thus named the coefficient of variation of the Rayleigh distribution $m$ after Motokawa.

Thus, if a signal is Gaussian noise, its envelope should have a \cve equal to $m$ independently of its power, as \cve $= m$ holds for all Rayleigh distributions regardless of the scale parameter $\sigma$, or equivalently, regardless of the variance $\sigma^2$ of the corresponding Gaussian distribution. In other words, \cve is a scale-independent parameter. It is important to underline that  \cve $= m$ is valid for wide-band Gaussian noise of arbitrary power as well as for any filtered sub-band \citep{schwartz1995communication}. Thus, the invariance of $m$ for Gaussian noise serves as the core mathematical result to test the Gaussianity of time series.

%% file: sections/3.NumericalResults.tex
\section{\cve for sampled Gaussian noise}

In this section we study the behavior of \cve in the discrete-time case. We performed Monte Carlo simulations and obtained the CVE distribution of zero-mean Gaussian noise.

\begin{figure}[t!]
    \centering
    \includegraphics[width=0.5\linewidth]{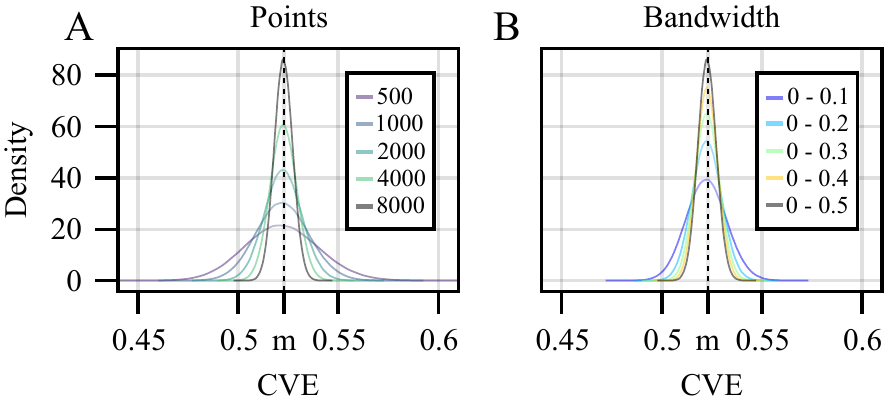}
    \caption{\cve for Gaussian noise of different number of points (A) and  bandwidth (B). Distributions 8000 and 0-0.5 are the same.}
    \label{fig1}
\end{figure}

We simulated $10^6$ instances of Gaussian noise epochs, calculated their \cve, and obtained their experimental distribution under different sample sizes and filtering bandwidths (\autoref{fig1}). Filters were low pass  with zero-phase response for reasons explained in Section 5. All distributions were unimodal with a mode around $m$, and their variance decreases for smaller sample sizes and narrower filtering. The \cve of Gaussian epochs under strong constraints tend to a mode lower than $m$, e.g. with very few samples (\autoref{fig1} left panel, wider distribution) or very narrow bandwidths (not shown).

Overall, the distribution for the \cve of discrete-time zero-mean Gaussian noise is a function of the number of points $N$ available and the bandwidth $BW$ of the signal, and computer simulations allow us to test the hypothesis of Gaussianity using \cve. Thus, as a first approach, the \cve of sampled Gaussian noise can be used to generate probability models and, for a given $\alpha$ value, create confidence intervals to test Gaussianity. 

\subsection{The case of Rician fading}
A multipath propagation channel with a dominant component deviates from Rayleigh fading and displays Rician fading instead. Interestingly, while the coefficient of variation of the Rician distribution is not unique, it can still be used to test for non-zero mean Gaussian noise.

The envelope of non-zero mean Gaussian noise follows a Rician distribution \citep{papoulis1989probability}:
\begin{equation}
    f_{Rice}(x;\nu,\sigma)=  \frac{x}{\sigma^2}e^{\frac{-(x^2+\nu^2)}{2\sigma^2}} I_{0}\left(\frac{x\nu}{\sigma^2}\right)
\end{equation}
where $I_{0}$ is the modified Bessel function of the first kind and order zero. The mean and standard deviation of the Rician distribution are
\begin{equation}
\begin{aligned}
  \mu &= \sigma\sqrt{\frac{\pi}{2}}L_{1/2}\left(\frac{-\nu^2}{2\sigma^2}\right), \\
  std &= 2\sigma^2 +\nu^2 \frac{\pi\sigma^2}{2}L^2_{1/2}\left(\frac{-\nu^2}{\sigma^2}\right) 
\end{aligned}
\end{equation}
\noindent where $L_{1/2}$ is a Laguerre polynomial.

\begin{figure}[t!]
    \centering
    \includegraphics[width=0.5\linewidth]{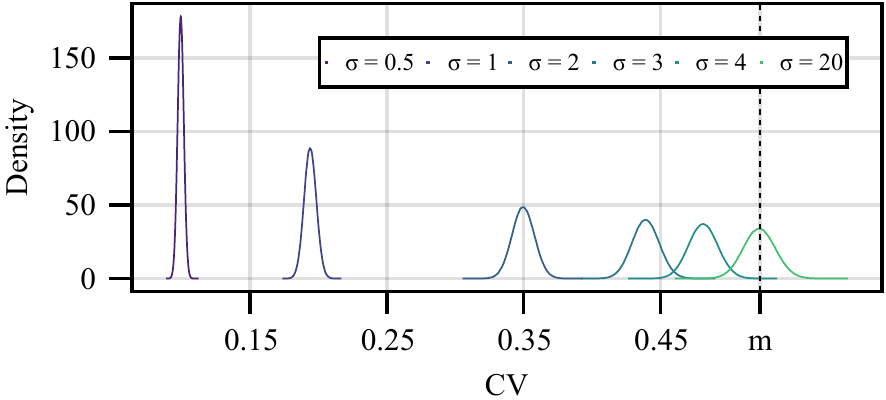}
     \caption{CV of the Rice distribution with $\nu=5$ and different $\sigma$.}
    \label{fig2}
\end{figure}

Thus, the coefficient of variation of the Rician distribution varies depending on the values of $\nu$ and $\sigma$. Nonetheless, Monte Carlo simulations show that for a given pair $\nu, \sigma$ the CV of the Rician distribution is fixed, and thus can also be used to test for Gaussianity in the Rician fading case (\autoref{fig2}). For Rician distributions with $\nu = 0$, it is well-known that the Rician distribution collapses into the Rayleigh distribution, and as expected the resultant CV is centered around $m$ (not shown). For $\nu \neq 0 $ the CV depends on the ratio $\nu/\sigma$(proportional to the ratio between the power of the dominant component and the multipath component) and approaches 0 for $\nu>>\sigma$ and $m$ for $\nu<<\sigma$ (\autoref{fig2}), showing that for small $\nu/\sigma$ the dominant signal is not dominant anymore, and we re-obtain a Rayleigh fading signal. 

\section{\cve characterizes amplitude modulation patterns}
\cve is also a measure of the amplitude modulation pattern of a signal. The \cve of Rayleigh fading identifies the characteristic amplitude modulation of Gaussian noise, while deviations from $m$ are connected to either rhythmic or pulsating modulation regimes.

\begin{figure}[h!]
    \centering
    \includegraphics[width=0.5\linewidth]{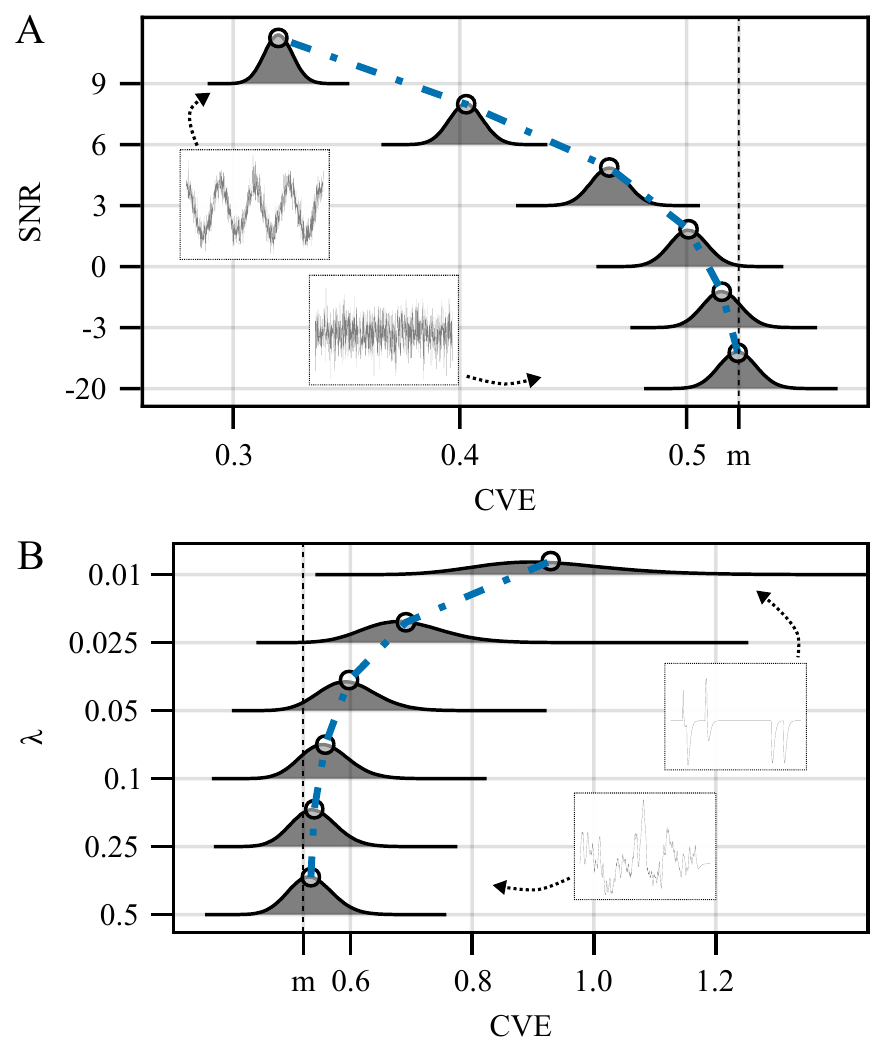}
    \caption{\cve for (A) rhythmic and (B) pulsating amplitude modulations. }
    \label{fig3}
\end{figure}

Sinusoidal signals exhibit a sub-Gaussian \cve, as the instantaneous amplitude of pure tones is constant, pushing the variance of the envelope and the \cve $\to 0$. For a sinusoid transmitted through a noisy channel, the signal-to-noise ratio (SNR) defines the amplitude modulation and the corresponding \cve distribution. When SNR is low, signals behave like Gaussian noise (i.e. \cve near $m$), but as the signal power increases the envelope becomes less noisy resulting in a sub-Gaussian \cve (\autoref{fig3} A).  These rather rhythmic modulation regimes are found in continuous wave radar, and they are also of interest in cognitive radio, where the non-Gaussianity of the transmission channel needs to be detected.

Pulsating modulations are also encountered in many applications such as those used in pulse radar or the energy modulation found in shot noise. These phasic amplitude modulations are easily detected using \cve as the increase in variance of the instantaneous amplitude results in a super-Gaussian \cve. For a filtered Poisson process, the \cve depends on the density (rate) of the elementary pulses. When densities are low, the corresponding CVE increases.  However, as the density of pulses increases, the overall signal becomes Gaussian and its CVE decreases towards $m$ (\autoref{fig3} B), recovering the well-known Gaussian limit of shot noise at high intensities.

\section{Fourier representation of amplitude modulation patterns}
\begin{figure}[h!]
    \centering
    \includegraphics[width=0.5\linewidth]{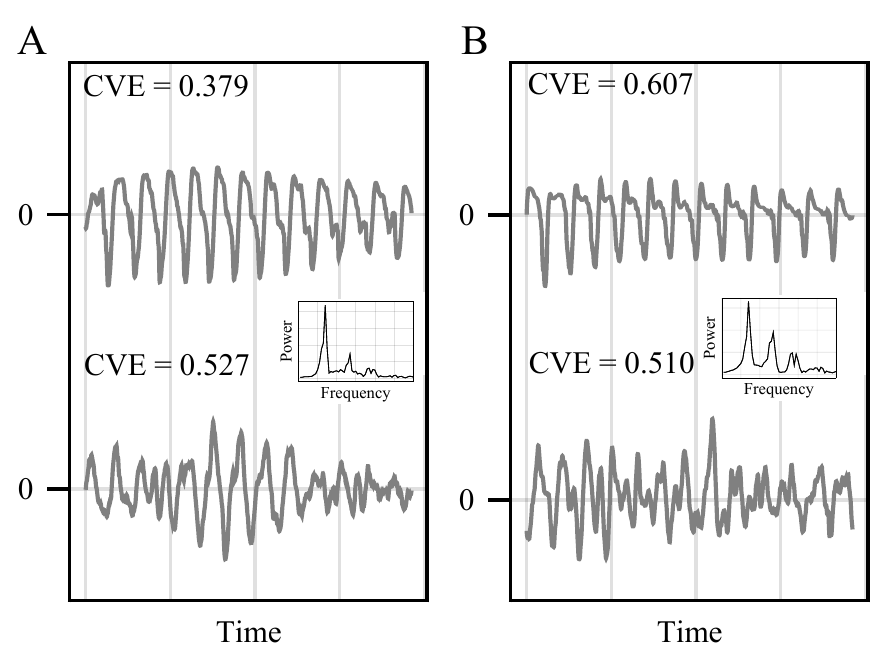}
    \caption{Experimental EEG data (upper traces) and one instance of their corresponding phase-randomized versions (lower traces). }
    \label{fig4}
\end{figure}

In many domains of signal processing, a simple calculation of the power spectrum of a signal is commonly used to indicate the power emitted in a small frequency sub-band. Typical uses of this approach can be found in EEG, where the energy in the $\alpha, \delta, \gamma$ bands are relevant clinical indicators or in seismology, where spectra are important to understand quake geodynamics. However, \cve opens a new avenue for signal analysis as two signals with identical power spectra can differ in their \cve and, hence, in their temporal profile. Thus, for example when signals representing the epoch of a neonate during convulsion  (\autoref{fig4} A and B, top traces) (data from \citep{stevenson2019dataset},  subject 44) are transformed using Fourier transform phase randomization (FTPR) \citep{theiler1992testing}, the original and phase-randomized signals have exactly the same power spectra, but interestingly their temporal (amplitude modulation) profiles differ along with their \cve (\autoref{fig4} A and B, lower traces). In the example of \autoref{fig4}, the original trace in A (upper trace) presents a rhythmic modulation and the original trace in B (upper trace) exhibits a pulsating modulation, but these regimes are disrupted by FTPR and transformed in Rayleigh fading profiles.

\begin{figure}[h!]
    \centering
    \includegraphics[width=0.5\linewidth]{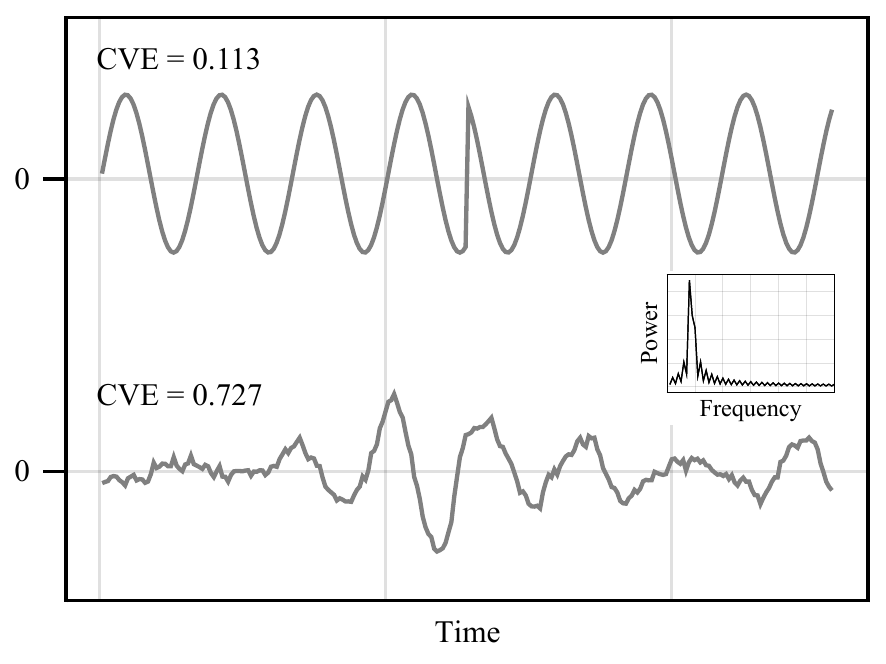}
    \caption{A simulated time series (upper trace) with one of its phase-randomized surrogates (lower trace).}
    \label{fig5}
\end{figure}

A more extreme case can be observed with an artificial signal made of a sine wave combined with a cosine wave of the same frequency. In \autoref{fig5}, the original signal (upper trace) has a \cve = 0.113 indicating it is almost a pure tone, but if we explore the space of its phase-randomized versions, it is possible to find a signal (lower trace) with the same power spectrum but a \cve = 0.727 similar to a deformed Gabor temporal pulse. Taking together the results from  \autoref{fig4} and \autoref{fig5}, it is clear that the \cve is a parameter that provides an overall measure of the temporal profile of a given signal independently of the power spectrum. Additionally, \cve is a measure of the structure of the phase profile of a signal --- information that is lost when using the spectral profile alone. Of course, this result is far from new. In effect, for any signal and any of its phase-randomized versions, their power spectra are identical but their Fourier transforms are not. Unfortunately, in many cases the complexities of Fourier analysis are subsumed into the simple evaluation of the power spectrum, thus discarding temporal profile variations \citep{galka2000topics}.  

\section{Estimator properties of CVE}

\begin{figure}[h!]
    \centering
    \includegraphics[width=0.5\linewidth]{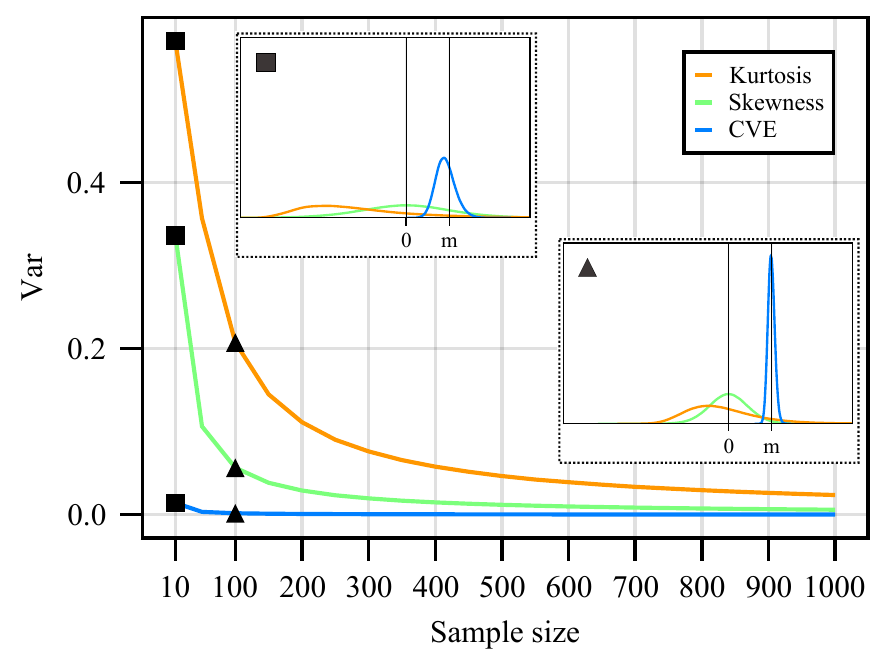}
    \caption{Estimator variance for kurtosis, skewness and \cve. Insets display the estimator distribution at the given sample size.}
    \label{fig6}
\end{figure}

\begin{figure}[h!]
    \centering
    \includegraphics[width=0.5\linewidth]{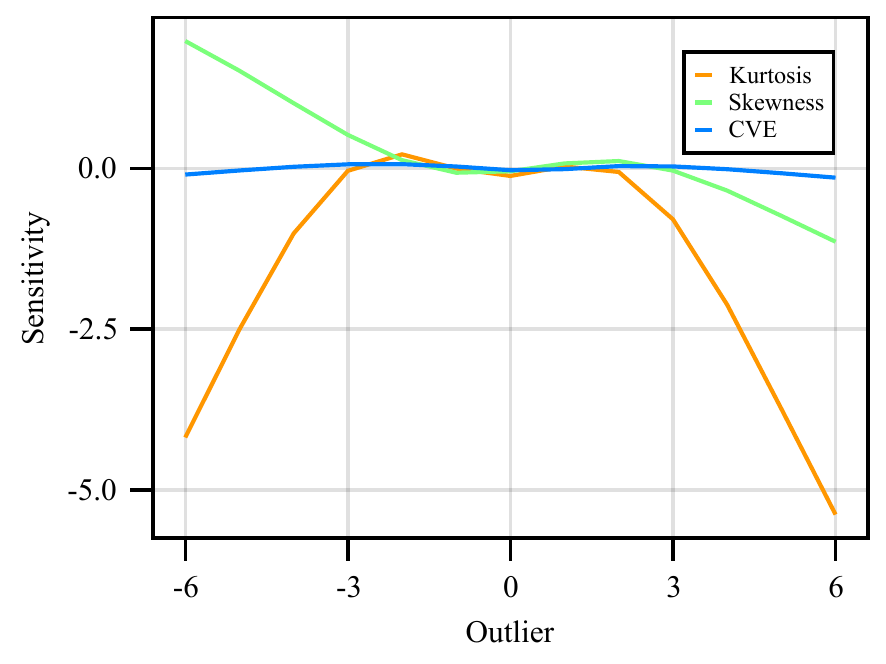}
    \caption{Sensitivity measured as the difference between the value of the estimator for 20 samples from $\mathcal{N}(0,1)$ and the value of the estimator for this same sample plus the outlier. The mean of $10^6$ calculations is plotted.}
    \label{fig7}
\end{figure}

The classic time-domain parameters for Gaussianity testing are skewness and (excess) kurtosis, and CVE offers some advantages over these statistics. CVE displays a faster rate of convergences to its continuous-time value than skewness and kurtosis (\autoref{fig6}). For \cve, estimator variance approaches 0 much faster than skewness and kurtosis, is significantly smaller even for small sample sizes, and remains smaller for large sample sizes as well. \cve is a consistent estimator and its estimator bias decreases much faster than the bias of kurtosis. In the frequency domain, \cve displays an equivalent performance (not shown). A simple sensitivity test indicates that \cve is less influenced by outliers than skewness and kurtosis (\autoref{fig7}). Also, \cve $=m$ is a unique property of Gaussian signals, but the uniqueness of the Gaussian skewness and kurtosis may not be guaranteed, so \cve is expected to outperform these estimators in these cases. These characteristics make \cve an interesting parameter for algorithms used to establish the Gaussianity of experimental time series.

%% file: sections/4.Discussion.tex
\section{Discussion}
We have presented an approach to assess Gaussianity based on considering signals as amplitude modulation channels, opening a new avenue for signal classification. The core of this approach depends on a property of the envelope of Gaussian noise:  the ratio $stdev/mean$ for the envelope of infinite Gaussian noise equals \meqgcveapprox. For discrete-time signals, it is possible to build empirical distributions that are centered on $m$ but whose dispersion depends on the number of points available and signal filtering. \cve allows precise confidence intervals to be constructed from these distributions according to the parameters of the experimental signal,  thus acting as a discriminating statistic --- essentially a fingerprint --- to test Gaussianity. Furthermore, \cve identifies the amplitude modulation of a signal. Sub-Gaussian \cve values reflect rhythmic signals, while super-Gaussian \cve values identify pulsating signals. Also, \cve is connected to the Fourier phase profile of the sampled epoch. Thus, two epochs with exactly the same power spectrum can have very different \cve values. Moreover, the use of \cve is simple to implement, more robust and not computationally intense. Indeed, it can be applied online as an automatic signal classifier because it does not require the adjustment of parameters once the sample size and filtering characteristics of data acquisition are defined.